\documentstyle[twoside,12pt]{article}
\begin{document}
\begin{titlepage}
\begin{flushleft}        
       \hfill                      USITP-99-06\\
      \hfill                      hep-th/9908001\\   	                                 
\end{flushleft}
\vspace*{3mm}
\begin{center}
{\LARGE The Hagedorn transition, Deconfinement\\ 
         and \( {\cal N}=4 \) SYM Theory\\}
\vspace*{12mm}
{\large Bo Sundborg\footnote{E-mail: bo@physto.se} \\
        {\em Institute of Theoretical Physics \\
        Box 6730\\
        S-113 85 Stockholm\\
        Sweden\/}\\}
\vspace*{25mm}
\end{center}

\begin{abstract} 
\( {\cal N}=4 \) Super Yang-Mills theory supplies us with a 
non-Abelian 4D gauge theory with a meaningful perturbation expansion, 
both in the UV and in the IR. We 
calculate the free energy on a 3-sphere and observe a deconfinement 
transition for large N at zero coupling. The same thermodynamic 
behaviour is found for a wide class of toy models, possibly also 
including the case of non-zero coupling. Below the transition we also find 
Hagedorn behaviour, which is identified with fluctuations signalling the 
approach to the deconfined phase. The Hagedorn and the deconfinement 
temperatures are identical. Application of the AdS/CFT correspondence gives a 
connection between string Hagedorn behaviour and black holes.
\end{abstract} 

\end{titlepage}

\section{Introduction}

In large \( N \) discussions of Yang-Mills theory the thermal 
deconfinement transition is accompanied by a jump in the free energy 
from order \( N^0 \) to order \( N^2 \). The confined phase consists 
of hadrons and glueballs, which should have spectra with narrow 
resonances. Models for such spectra like statistical bootstrap models 
or string theory, often have densities of states rising exponentially 
with energy, i.e. Hagedorn behaviour \cite{Hagedorn}. Hagedorn 
spectra themselves give rise to singular thermodynamics at a 
characteristic temperature because Boltzmann suppression of high 
energy states can be compensated by the sheer numbers of states. 
It would be natural if this Hagedorn temperature and the 
deconfinement temperature were the same. Thorn \cite{Thorn:1981iv} has 
discussed possible alternatives but it seems that the question for
the generic 4-dimensional gauge theory is too difficult to handle. 
Pisarski has also argued that a second order deconfinement transition 
at infinite \( N \) implies a Hagedorn spectrum \cite{Pisarski:1984db}.

In string theory, which has Hagedorn behavior at zero coupling, this supposed 
relation has even been used to
model what might happen above the Hagedorn temperature 
\cite{Atick:1988si}, which for closed 
string theories is difficult to interpret as a true limiting temperature,
since the energy density never diverges \cite{Sundborg:1985uk}.
Furthermore, the vanishing of an effective temperature-dependent string tension at 
the Hagedorn temperature has been naturally associated with deconfinement 
\cite{Pisarski:1982cn,Olesen:1985ej,Salomonson:1986eq}. Still, until 
now explicit calculations covering the full range below, close to, and above
the transition has been 
missing. To actually demonstrate that the Hagedorn and 
the deconfinement temperatures are identical, and that the thermodynamics
below this critical temperature can equally well be described by an 
exponential density of states or some mechanism binding otherwise free 
colour charges, requires large computational power or very simple 
models, as those studied in the present paper.
 
\( {\cal N}=4 \) Super-Yang-Mills at zero coupling on a 
three-sphere is simple enough for explicit calculations, but is also 
smoothly related to the interacting theory. The dimensionless coupling 
is an actual 
parameter of the theory because of conformal invariance. Although we 
perform an explicit calculation for this theory most of the 
calculations of 
this paper are phrased in terms of a much more general class of toy 
models, including free gauge theories 
with adjoint matter on spheres. For asymptotically free theories our 
calculation should be directly relevant in the limit of small radius 
(implying small coupling). Note that zero gauge coupling is understood 
in this paper as meaning free dynamics, but with a global colour neutrality 
constraint on the states. The rationale for this constraint is that it 
is present for any nonzero coupling due to Gauss law on a compact 
manifold. Thus it ought to be kept in taking a smooth free limit.

The \( {\cal N}=4 \) theory is particularly interesting because the 
AdS/CFT correspondence 
\cite{Maldacena:1997re,Gubser:1998bc,Witten:1998qj,Aharony:1999ti} can 
be used to discuss gravitational and 
string theory interpretations of all SYM results, cf. sect. 
(\ref{AdS/CFT}). For strong coupling gauge theory this correspondence 
has been used by Witten \cite{Witten:1998qj,Witten:1998zw} 
to extract information about gauge theory deconfinement from a 
gravitational phase transition found by Hawking and Page \cite{Hawking:1983dh}. 
High and low temperature limits at weak coupling, but 
not intermediate temperatures, are also
discussed in \cite{Witten:1998zw}. Corrections to the strong coupling 
results on the sphere have been studied in 
\cite{Gao:1998ww,Landsteiner:1999gb,Caldarelli:1999ar}. On the basis 
of these corrections it has been claimed that the Hawking-Page phase 
transition disappears at small coupling 
\cite{Gao:1998ww,Caldarelli:1999ar}, whereas the result of the 
present paper is that there is a similar phase transition also at zero 
coupling. A possible source of the discrepancy 
is that these authors consider string 
corrections to the action and the metric, but not the effects of string spectra on 
the thermodynamics, which by the Maldacena conjecture ought to be automatically 
represented by the 
gauge theory calculation. In contrast, the high temperature expansion considered
in \cite{Burgess:1999vb} seems to be consistent with the similarity 
between strong and weak coupling results which we find. The 
implications of the AdS/CFT correspondence for the relation between 
the Hagedorn transition and deconfinement has been investigated in a 
different way in \cite{Rama:1998cb}.

We always 
study the large \( N \) limit of gauge theories with 
constituent states in the adjoint representation of the gauge group.  
The other essential 
ingredients are confinement, and a discrete energy spectrum with a 
positive energy ground 
state for the constituents. There are no interactions 
among the colour charges in the model except for a singlet constraint 
from confinement. This makes the model exactly calculable. Hopefully the 
results are stable to the introduction of small interaction terms. The 
models we are considering are 
slight generalizations of singlet ideal gases in flat space 
studied by Skagerstam \cite{Skagerstam:1984gv}, and two-dimensional lattice gauge theories with a 
Wilson action in the adjoint representation. The crucial 
difference that makes us able to directly identify the Hagedorn behaviour 
is that we include the fluctuations around the leading large \( N \) 
result below the phase transition.

An important feature in our model is that deconfinement 
never has to take place on the level of states, i.e. no non-singlets 
have to be taken into account above the transition. In string 
language there is no ``string breaking'' above the Hagedorn transition. 
Instead, it seems that a finite N leads to a density of states that is 
only approximately exponential, so that a sufficiently high temperature 
discloses a number of constituent degrees of freedom which is not 
infinite but proportional to \( N^2-1 \). There appears to be a string exclusion 
principle \cite{Maldacena:1998bw} at work. Perhaps this is a more general 
property of ``deconfinement''. In any case, our models are explicit 
examples demonstrating how thermodynamics of colour neutral states may 
display ``deconfined'' properties.

The paper is organized as follows. In section (\ref{Csto}) the problem of counting 
singlet states is formulated and the solution is found for general spectra of 
constituents. In section (\ref{Ht}) it is demonstrated that Hagedorn 
behaviour results from these state spaces\footnote{After completing 
this work I realized that similar but simpler toy models giving 
Hagedorn behaviour are discussed by Gao and Li \cite{Gao:1998ww}. 
However, in their simplified models it is impossible to find the relation 
to a first order large \( N \) transition.}, and the grand canonical 
partition function close to the Hagedorn temperature is estimated. 
Partition sums for our prime example, \( {\cal N}=4 \) SYM on \( S^3 \), 
are calculated in section (\ref{N=4}). The details of the \( {\cal N}=4 \) results are 
not used elsewhere in the paper, but since we have not found them in 
the literature, we write them down explicitly. The
next section (\ref{sc}) deals with the problem of directly projecting 
the full grand canonical partition function to the physical singlet 
states in the large \( N \) limit. The technique we use is known 
before (see for example \cite{Skagerstam:1984gv}), and an approximate leading 
order result can almost be taken literally from this reference. By also 
pointing out the role of large \( N \) fluctuations we can 
however establish our main new result: In the systems under study, the 
deconfinement transition is accompanied by Hagedorn behaviour with 
the same critical temperature. Then in section (\ref{Z}) there is a short 
discussion of 
the role of unbroken symmetry under the center of the gauge group 
\(Z_{N}\). In section (\ref{AdS/CFT}) connections are made with 
gravitational physics and we argue that the similarity with between 
the present weak coupling results and Witten's results at strong 
coupling \cite{Witten:1998qj,Witten:1998zw} indicates that one could also discuss 
black holes in the AdS dual of the weak coupling theory. It is also 
argued that AdS black hole thermodynamics is related to Hagedorn 
thermodynamics. Finally in section (\ref{Conclusions}) we state our 
conclusions.
 
\section{Counting single trace operators}
\label{Csto}

Let us assume that we have constituent degrees of freedom with discrete 
energy spectra. Temporarily disregarding the colour degrees of freedom 
and the singlet condition 
we can write down partition sums \( \zeta_B \) and \( \zeta_F \) for 
constituent bosons and fermions, in terms of the energies \( E_{B,n} \) 
and \( E_{F,n} \) for all their states,
\begin{eqnarray}
\zeta_B(x)&=&\sum\limits_{n=1} {x^{E_{B,n}}},
	\label{CBosons} \\
\zeta_F(x)&=&\sum\limits_{n=1} {x^{E_{F,n}}},
	\label{CFermions} \\
\zeta(x)&=&\zeta_B(x) + \zeta_F(x)
	\label{CAll}
\end{eqnarray}
where we have defined
\begin{equation}
x\equiv e^{-\beta }\equiv e^{-1/T}.
	\label{xTDef}
\end{equation}
If we then assume that these fields are in the adjoint representation 
of \( SU(N) \), they can be represented by traceless hermitean matrices, 
and all gauge invariant states can be written as products of traces of 
hermitean matrices, each labelled by one of the discrete constituent 
states. The single-trace states
\begin{equation}
\left| {i_1,i_2,\ldots ,i_n} \right\rangle=
{\textnormal{Tr} } (\phi _{i_1}\phi _{i_2}\ldots \phi _{i_n})
\left| {i_1} \right\rangle\left| {i_2} \right\rangle\ldots 
\left| {i_n} \right\rangle
	\label{States}
\end{equation}
play the role of bound states, with the singlet 
condition being the only interaction we consider. They are 
bosons or fermions depending on whether they contain an even or odd 
number of constituent fermions. Because of the trace, cyclical 
permutations of the  constituent-state labels give rise to the same bound 
state. The cyclicity may be taken into account 
by thinking of bound states as ``necklaces'' of constituent ``beads''. For 
finite groups there are in general many relations between the traces. 
In case of a single matrix, \( \mathrm{Tr}(A-\lambda_1) (A-\lambda_2) 
\ldots (A-\lambda_{N-1})  =0 \) gives such a 
relation since sums of powers of the eigenvalues \( \lambda_i \) are 
traces of powers of \( A \). In the large \( N \) limit of \( SU(N) \) 
only two basic kinds of relations survive: \( \textnormal{Tr} A=0 \) for 
any constituent state \( A \) due to tracelessness, and 
\( \textnormal{Tr} \psi^2 =0 \) from the Pauli principle for fermionic 
constituents \( \psi \). The reason that the Pauli principle is effective 
for squared constituents but not in other products involving fermions is 
that only the trace of the square forces a multiplication of identical 
fermion Lie algebra components. In other cases the fermions can be in 
different Lie algebra directions.

The bound state partition sum \( z(x) \) may now be obtained by counting necklaces 
of constituent states, 
weighting each energy level with the corresponding Boltzmann weight 
\( x \), and finally subtracting the constituent partition sum and the 
sum for two identical fermion constituents. Fortunately, the necklace problem is a 
special case of a class of 
combinatorial problems solved by P\'{o}lya's theorem \cite{Polya}. The number 
of bound states with \( n \) constituents can be found from the partition sum
\begin{equation}
{1 \over n}\sum\limits_{k\left| n \right.} 
{\varphi (k)(\zeta _B(x^k)}+\zeta _F(x^k))^{n / k},\quad n>2,
	\label{PolyaSum}
\end{equation}
over divisors of \( n \), where \( \varphi \) is called the Euler totient 
function. \( \varphi(n) \) is 
the number of positive integers less than \( n \) which are relatively 
prime to \( n \), 
with \( \varphi(1)=1 \) by definition. Summing over numbers of constituents 
we obtain
\begin{eqnarray}
		z(x)&=-(\zeta _B(x)+\zeta _F(x))-{1 \over 2}\zeta _F(x^2)+
\sum\limits_{n=1}^\infty  {{1 \over n}\sum\limits_{k\left| n \right.} 
{\varphi (k)(\zeta _B(x^k)}+\zeta _F(x^k))^{n / k}}
	 \\
	&=-\zeta _B(x)-\zeta _F(x)-{1 \over 2}\zeta _F(x^2)-\sum\limits_{k=1}^\infty  
  {{{\varphi (k)} \over k}}\log \left[ {1-\zeta _B(x^k)-\zeta _F(x^k)} 
  \right],\label{SingleTraceSum}
\end{eqnarray}
where the order of summation has been changed and the sum over 
products of \( k \) has been performed in the last line. The first 
three negative terms are there to cancel unphysical traces of single 
operators and two identical fermions.

\section{The Hagedorn transition}
\label{Ht}

One may see quite easily that the number of states grows exponentially 
with energy if there are at least two constituent states. For \( n \) the 
number of factors inside a trace, the cyclicity relation becomes less and 
less important for large \( n \) and can at most give a factor \( 1/n \). 
Neglecting this factor the number of states involving only the two 
lowest energy constituents grows at least as \( 2^n \), 
and the energy is at most \( n E_2 \), where \( E_2 \) is the second 
lowest energy. Therefore the density of states grows at least as \( 
2^{E/E_2} \).

More precisely, the partition sums (\ref{SingleTraceSum}) have lowest 
temperature singularities at
\begin{equation}
1=\zeta (x_H) \equiv \zeta _B(x_H)+\zeta _F(x_H).
	\label{Hagedorn}
\end{equation}
Close to this point we may approximate
\begin{equation}
z(x)\approx -\log [1-\zeta (x)] \approx -\log [(x_H-x)\zeta '(x_H)],
	\label{SingPartSum}
\end{equation}
and from
\begin{equation}
z(x) =  \sum\limits_{n=1}^\infty d(n) x^n
	\label{degeneracy}
\end{equation}
and residue calculus we obtain the contribution to the asymptotic level degeneracy
\begin{eqnarray}
	 d(n)&=&
	{i \over {2\pi}}\oint {{{dx} \over {x^{n+1}}}}\log [1-\zeta (x)]
	\approx {i \over {2\pi }}\oint {{{dx} \over {x^{n+1}}}}
	\log [(x_H-x)\zeta '(x_H)]
	 \\
  &\approx&-{{x_H^{-n}} \over {2\pi i}}\oint {{{dy} \over {y^{n+1}}}}\log [1-y]
  	\label{AsymDeg}
\end{eqnarray}
in case the spectrum is integrally spaced. The asymptotic 
density of states should have the same Hagedorn behaviour at \( x=x_H \) 
irrespective of this simplifying assumption. 

Because there are so many different states that are easily excited 
close to the Hagedorn temperature, the effects of identical particles 
in Bose or Fermi statistics vanishes for the singular behaviour at this 
temperature. Using Boltzmann statistics we then find the 
the grand canonical partition function 
\begin{equation}
Z(x)\approx \sum\limits_{n=1}^\infty  {{1 \over {n!}}}z(x)^n=e^{z(x)}
\approx {1 \over {(x_H-x)\zeta '(x_H)}}
	\label{CanonicalHagedorn}
\end{equation}
close to \( T_H \).

\section{The \( {\cal N}=4 \) Super Yang-Mills partition sum on \( S^3 \)}
\label{N=4}

The results above are quite general, and the arguments in section 
(\ref{sc}) establishing a connection between Hagedorn behaviour and 
deconfinement in small volumes do not depend on the details of the 
partition sums. Nevertheless, because of the special importance of \( 
{\cal N}=4 \) SYM we include in this section an explicit calculation of 
its free partion 
sums on a sphere, which we have not found elsewhere in the 
literature. Readers who are not interested in these details may skip 
this section. 

Due to conformal invariance it is simple to find the
spectrum of \( {\cal N}=4 \) SYM on a three-sphere at zero coupling. There 
are ambiguities in how 
a field theory should be coupled to a background curvature, although the 
leading long distance behaviour is fixed by the equivalence principle. 
For SYM, the ambiguity may be fixed by asking that 
scalars are 
conformally coupled. 

Then one may take over results from flat spacetime to \(S^3 \times 
\mathbf{R}\)  by a conformal mapping \cite{Witten:1998zw}. The generator of scale 
transformations in 
the flat Euclidean theory is mapped to the Hamiltonian on the sphere. 
Because 
these operators are also related by an isomorphism of the conformal 
group they have identical spectra. We can conclude that energies of 
states on 
the sphere (of unit radius) are given by scaling dimensions of operators 
in flat spacetime. (In two dimensions this is the mapping between the 
plane and the cylinder.) 

On the sphere, all states have to be gauge singlets because of the Gauss 
law constraint and the lack of a boundary. We can then use the 
general procedure above to find the bound state partition sum and the 
grand canonical partition function close to the Hagedorn temperature. We 
only need the constituent Bose and Fermi partition sums. The constituent 
fields, scalars, fermions and gluons, their respective scaling 
dimensions and their partition sums are given in
\begin{equation}
\matrix{\textstyle{Field}&\textstyle{Dimension}&(1-x)^4  \zeta(x) \cr
\Phi &1&{{{6x} }-{{6x^3} }}\cr
\lambda &{{\textstyle{3 \over 2}}}&{{{8x^{3 / 2}} -8x^{5 / 2}}}\cr
{\bar \lambda }&{{\textstyle{3 \over 2}}}&{{{8x^{3 / 2}} -8x^{5 / 2}}}\cr
F&2&{{{6x^2} }-{{8x^3} +6x^4}}\cr
}
	\label{ConstituentSums}	
\end{equation}
where the factor \( (1-x)^{-4} \) in \( \zeta(x) \) arises from counting possible 
derivative operators (conformal descendants). Terms corresponding to 
equations of motion and their derivatives have been subtracted, since 
they give rise to trivial correlation functions\footnote{One may check, 
most simply for a conformal scalar on \( S^2 \), that this subtraction 
gives the correct counting of states.}. The gluon sum is 
somewhat more complicated and we write it down explicitly to 
demonstrate the procedure:
\begin{equation}
{{6x} \over {(1-x)^4}}-{{2x^4} \over {(1-x)^4}}-{{8x^3} \over 
{(1-x)^3}}.
	\label{GluonSums}	
\end{equation}
The three terms signify derivates of the six components of the 
field strength, a subtraction of derivatives of \( 
\partial^{\mu}\partial^{\nu} F_{\mu\nu} \) and 
\( \partial^{\mu}\partial^{\nu \:*\!}F_{\mu\nu} \) vanishing by 
antisymmetry, and finally a subtraction of derivatives of the 
equations of motions and the Bianchi identities (note from the 
denominator of the last term that it effectively includes derivatives 
in only three directions, so as not to subtract again terms 
vanishing automatically by antisymmetry). We are 
now ready to write down 
the partition sums for free \( {\cal N}=4 \) SYM at the limit 
\( N=\infty \) :
\begin{eqnarray}
	\zeta^{SYM}_B(x)&=&{{6x+6x^2-14x^3+6x^4} \over {(1-x)^4}}
	\label{SYMConstBosons} \\
	\zeta^{SYM}_F(x)&=&{{16x^{3/ 2}-16x^{5/ 2}} \over {(1-x)^4}}
	\label{SYMConstFermions} \\
	z^{SYM}_B(x)&=&-\zeta^{SYM}_B(x)-{1 \over 2}\zeta^{SYM}_F(x^2) \\
	&-&\sum\limits_{k=1}^\infty  {{{\varphi (k)} \over 2 k}
	\log \left[ {1-2\zeta^{SYM}_B(x^k)+\zeta^{SYM}_B(x^k)^2- 
	\zeta^{SYM}_F(x^k)^2} \right]}
    \label{SYMBosons} \\
	z^{SYM}_F(x)&=&-\zeta^{SYM}_F(x)-\sum\limits_{k=1}^\infty  
	{{{\varphi (k)} \over 2 k}
	\log \left[ {{{1-\zeta^{SYM}_B(x^k)-\zeta^{SYM}_F(x^k)} \over 
	{1-\zeta^{SYM}_B(x^k)+\zeta^{SYM}_F(x^k)}}} \right]}.
		\label{SYMFermions}
\end{eqnarray}
The Bose and Fermi partition sums are obtained by keeping track of 
whether the states contain an even or odd number of constituent 
fermions. Note that the sum of these terms reproduces eq. 
(\ref{SingleTraceSum}). The Hagedorn temperature is then given by the lowest temperature 
singularity, which can be found by solving eq. (\ref{Hagedorn}). 
We find that \( x_{H} \approx 0.072 \). Furthermore,
the grand canonical partition function approaches (\ref{CanonicalHagedorn})
close to \( T_H \).

All these calculations were performed at zero coupling, but one may 
hope that our general toy models also capture some of the essential 
ingredients of the interacting case, perhaps by  renormalization of 
the constituent spectra.

\section{The ideal gas with a singlet condition}
\label{sc}

The scale of \( T_H \) is set by \( 1/R \), the inverse radius of the 
sphere, which is the only fundamental energy scale in the problem. Thus 
\( T_H \rightarrow 0\) in the flat space limit, and we seem to get a 
divergent free energy at any temperature! The argument 
for considering only gauge singlet states breaks down in a 
non-compact 
space, but on the other hand a singlet condition ought to make little 
difference in infinite volume. And we do not really want to consider 
non-singlet states in a non-Abelian gauge theory. At best they are going 
to be extremely sensitive to turning on the interactions. 

Instead, the problem is due to setting \( N=\infty \). The free energy at 
zero coupling 
in flat space is well known, and is proportional 
to \( N^2 -1 \). To go above the Hagedorn temperature we have to 
take the large \( N \) limit in a way that allows us to extract divergent 
factors of \( N^2  \). We should also compute quantities like the free 
energy, which scale simply with \( N \), rather than the bound state 
partition sum, which certainly changes with \( N \) due to trace 
identities, but in a much less regular fashion. 

We write down the grand canonical partition function with a singlet 
constraint as
\begin{equation}
\int\limits_{SU(N)} {dg}
\exp \left( {\sum\limits_{n=1}^\infty  {{{\zeta_B(x^n)} \over n}\chi (g^n)}-
\sum\limits_{n=1}^\infty  {(-1)^n{{\zeta_F(x^n)} \over n}\chi (g^n)}} \right)
	\label{SingletPartition}
\end{equation}
where \( g \) are elements of \( SU(N) \) and \( \chi(g) \) is the 
character of the adjoint representation. 
This equation can be derived 
in a general way by a coherent state technique, as in \cite{Skagerstam:1984gv}. 
Alternatively, one may check that the expression 
represents the proper combinatorics, by noting that singlets can be identified 
by acting with \( SU(N) \) transformations on the constituent states. It 
is enough to keep track of the eigenvalues \( R_{i}(g) \) of the representation 
matrices \( R(g) \). The rotated partition sum for bosons
\begin{eqnarray}
	&\prod\limits_k {\prod\limits_{i=1} 
(1+x^{E_k}R_i(g)+x^{2E_k}R_i(g)^2+\ldots )} \\
	&=\prod\limits_k {Det(1-x^{E_k}R(g))^{-1}}
  =\exp \left( {\sum\limits_k {\mathrm{Tr} \left( {\sum\limits_{n=1}^\infty  
  {{{x^{nE_k}} \over n}R(g^n)}} \right)}} \right) 
\end{eqnarray}
can be projected to singlet states by integration over the group, using 
the orthogonality properties of group characters \( \chi(g) \). Taking also 
the fermion contribution into account yields eq. (\ref{SingletPartition}).

In terms of the eigenvalues \( \exp(i\alpha_i) \) of \( g \) the adjoint 
character
\begin{equation}
\chi (g)=-1+
\left( {\sum\limits_{m=1}^N {e^{i\alpha _m}}} \right)
\left( {\sum\limits_{n=1}^N {e^{-i\alpha _n}}} \right)
=N-1+2\sum\limits_{m<n} {\cos (\alpha _m-\alpha _n)}.
	\label{AdjChar}
\end{equation}
The fact that the integrand only depends on the eigenvalues simplifies 
the group integral to an integral over these eigenvalues 
\begin{eqnarray}
	 &{1 \over {N!}}\int {\prod\limits_{i=1}^N {{{d\alpha _i} \over {2\pi }}}}
\prod\limits_{i\ne j} {\left| {2\sin \left( {{{\alpha _i-\alpha _j} \over 2}} 
\right)} \right|}
\left( {\sum\limits_{k=1}^N 
{\delta \left( {-{2\pi k} +\sum\limits_i {\alpha _i}} \right)}} \right) \\
	&\times\exp \left( {\sum\limits_{n=1}^\infty  
	{{1  \over n}{\left(\zeta_B(x^n)-(-1)^n\zeta_F(x^n)\right)} 
  \left( 
  {-1+\sum\limits_{i,j} {\cos \left( {n(\alpha _i-\alpha _j)} 
  \right)}} \right)}} \right).
		\label{NSingletPartition}
\end{eqnarray}
The terms with \( k \neq 0\) in the sum of \(\delta \) functions encode
that we are dealing with 
\( SU(N) \) rather than \( SU(N)/Z_N \). By 
standard large \( N \) techniques the discrete set of eigenvalues may be 
replaced by an eigenvalue density \( \rho \) such that
\begin{eqnarray}
	\rho(\alpha) &=&  {1 \over N}{dn \over d\alpha_n}
	\label{EigenvalueDensity}, \\
1&=&\int {\rho (\alpha )d\alpha }.
	\label{EigenvalueIntegral}
\end{eqnarray}
Since different orderings of the eigenvalues are related by Weyl 
reflections it is enough to consider a single ordering, which may be 
chosen increasing, thus making \( \rho \) positive. In the large \( N \) 
limit the \(\delta \) function in (\ref{NSingletPartition}) becomes 
irrelevant unless we are dealing with sharply peaked eigenvalue 
distributions which occur for very high temperatures. For the time being 
we can neglect this factor. Then we find a partition function
\begin{equation}
Z(x)\propto \int {D\rho \,e^{-S[\rho ]}},
	\label{NPartitionFunction}
\end{equation}
where
\begin{eqnarray}
	 &S[\rho ]=-N^2\int \!\!d\alpha\! \int\!\! d\beta \,
\rho (\alpha )K_{x}(\alpha ,\beta)\rho (\beta )
+\sum\limits_{n=1}^\infty  {1  \over 
n}{\left(\zeta_B(x^n)-(-1)^n\zeta_F(x^n)\right)}, \\
&K_x(\alpha ,\beta )\equiv \log \left| {2\sin \left( {{{\alpha -\beta } \over 2}}
 \right)} \right|+\sum\limits_{n=1}^\infty  
 {{1  \over n}{\left(\zeta_B(x^n)-(-1)^n\zeta_F(x^n)\right)}
 \cos \left( {n(\alpha -\beta )} \right)}
	\label{EffectiveAction}
\end{eqnarray}
can be regarded as an effective action for the eigenvalue density.
The stationarity condition in a steepest decent estimate for \( Z \)
then gives the integral equation
\begin{equation}
0=\mathrm{p.v.} \int {d\beta }
  \left\{ {{\textstyle{1 \over 2}}\cot \left( {{{\alpha -\beta } \over 2}}
   \right)-\sum\limits_{n=1}^\infty  {\left( {\zeta_B(x^n)-(-1)^n\zeta_F(x^n)} 
   \right)\sin \left( {n(\alpha -\beta )} \right)}} \right\}\rho (\beta ) 
   \label{IntegralEquation}
\end{equation}
for \( \rho \). (One does not 
get the integral of this equation since \( \rho \) satisfies an 
additional integral constraint.) Since the equation is translationally 
invariant in \( \alpha \)-space 
\( \rho \equiv 1/{2\pi} \) is always a solution, but there may be others with 
lower free energy, which thus would appear to break the translational 
symmetry spontaneously. As we shall see below, this is almost what happens for 
high enough temperature, but 
not quite. 

To identify the transition point where non-trivial solutions 
appear, and to prepare for our identification of Hagedorn behaviour, it is useful to view the 
effective action as a quadratic expression in 
\( \rho \). We can regard 
\begin{equation}
K_x:\rho (\alpha )\to K_x \rho (\alpha )
\equiv -\int {d\beta \,K_x(\alpha ,\beta )}\rho (\beta )
	\label{KOperator}
\end{equation}
as a symmetric operator. Then, modulo the additional conditions on \( \rho \),  
the search for the minimum of the free energy becomes the search for the 
lowest eigenvalue of \( K_x \).
There will be a transition point at a value of the temperature where the 
lowest \( K_x \)-eigenvalue becomes degenerate. Such a point is easy to find by 
Fourier expanding the \( K_x \)-eigenvalue equation. The \( K_x \)-eigenfunctions 
are found to be trigonometric functions:
\begin{eqnarray}
	K_x {e^{in\alpha }} &
=&{\pi \over n}\left( {1-\zeta_B(x^n)+(-1)^n\zeta_F(x^n)} \right)\;e^{in\alpha },
\quad n\ne 0 \\
	K_x 1 &=&0
	\label{Eigenvalues}
\end{eqnarray}
We find a transition point at an \( x=x_D=x_H \) satisfying
equation (\ref{Hagedorn}) determining the Hagedorn temperature.

The \( K_x \)-eigenfunctions which become degenerate with the constant solution 
indicate the nature of the transition. We have to require 
\( \rho \geq 0 \), as \( \rho \) is an
eigenvalue density, which should also integrate to \( 1 \). Therefore, 
the new \( K_x \)-eigenfunctions are not acceptable as such. Instead, the linear 
combinations
\begin{equation}
\rho (\alpha )={1 \over 2\pi}\left(1+a_1\cos (\alpha )+a_2\sin (\alpha )\right),
\quad  {a_1^2+a_2^2}\le 1
	\label{DegenerateSolutions}
\end{equation}
are the minimizing solutions at the transition 
point. At the boundary of the solution set we find the solutions \( \rho 
(\alpha )=\sin ^2(\alpha +\delta )/\pi\), which are closest to the new 
\( K_x \)-eigenfunctions. Apparently, at the 
transition it becomes 
energetically possible for the density to have a zero. On the other side 
of the transition the density can still not 
become negative, but the eigenvalue distribution can have ``gaps'', where 
the density vanishes. To solve this case, which involves new boundary 
conditions, a more direct approach to the integral equation (eq. 
\ref{IntegralEquation}) is needed.

This kind of boundary problem is commonplace in large \( N \) 
calculations \cite{Brezin:1978sv,Gross:1980he}. A somewhat unusual 
feature of equation 
(\ref{IntegralEquation}) is that the two-eigenvalue interaction is 
non-trivial, whereas a non-trivial single-eigenvalue potential is more 
familiar. Fortunately, there is still a mathematical theory for this kind 
of problem \cite{Mush}. Essentially, the mathematical 
method consists in multiplying the integral equation 
(\ref{IntegralEquation}) with an operator that is inverse to the first, 
singular term, in order to get a regular equation of Fredholm type. This 
equation either lacks solutions or has a finite-dimensional solution 
space, which may be searched for solutions to the original equation.

The higher, \( n>1 \),
Fourier components of the non-singular term in 
the integral equation (\ref{IntegralEquation}) are exponentially small 
relative to the \( n=1 \) term  at
finite temperatures, and go as inverse powers of \( n \) for 
asymptotically high temperatures, at least when the constituent density of 
states is asymptotically constant or increasing\footnote{This is the case 
for free field theory in one or more spatial dimensions.}. It is 
then reasonable to neglect the higher terms and study the approximate 
equations obtained by replacing the kernel \( K_x(\alpha ,\beta ) \) with
\begin{equation}
	K_x^1(\alpha ,\beta )
	\equiv \log \left| {2\sin \left( {{{\alpha -\beta } \over 2}} \right)} \right|
	+\left( {\zeta_B(x)+\zeta_F(x)} \right)\cos \left( {\alpha -\beta } \right).
		\label{ApproximateKernel}
\end{equation}

Precisely this mathematical problem is solved in \cite{Skagerstam:1984gv}. We 
find the same solution, but also note a 
misprint in eq. (19) of 
that reference. Instead, the free energy \( F=-\log Z \) above the transition reads
\begin{eqnarray}
F_>
\approx -{{N^2} \over 2}\left( {\zeta(x)-1+\sqrt {\zeta(x)^2-\zeta(x)}-
\log \left( {\zeta(x)+\sqrt {\zeta(x)^2-\zeta(x)}} \right)} \right)+\zeta(x)
	\label{FreeE}
\end{eqnarray}
We also give the solutions for the eigenvalue densities
\begin{eqnarray}
\rho_{\delta}(\alpha )&=&\rho_{0}(\alpha-\delta) \\
\rho_{0}(\alpha) &=&
\left\{ {\matrix{{
{{\cos ({\textstyle{1 \over 2}}\alpha )} \over {\pi \sin ^2({\textstyle{1 \over 2}}
\alpha _c)}}\sqrt {\sin ^2({\textstyle{1 \over 2}}\alpha _c)-
\sin ^2({\textstyle{1 \over 2}}\alpha )},\quad\left| \alpha  
\right|<\alpha _c}\qquad \cr \cr
{0,\quad\left| \alpha  \right|>\alpha _c}
}} \right.\\
  \sin ^2({\textstyle{1 \over 2}}\alpha _c)&=&1-\sqrt {1-{1 \over {\zeta(x)}}}
  \label{EigenvalueD}
\end{eqnarray}
At the transition, these solutions to an approximate problem agrees with the 
extreme cases of the exact solutions (\ref{DegenerateSolutions}) to the 
exact problem. We do however only expect qualitative and approximate 
agreement above the transition. If we calculate the effective action 
(\ref{EffectiveAction}) below the transition we can also estimate the free 
energy 
\begin{equation}
	F_<=\zeta(x)
	\label{FreeE<}
\end{equation}
which together with eqs. (\ref{FreeE}) signifies a first 
order large \( N \) transition at the Hagedorn temperature. 

The \( N^2 \) dependence of the high temperature phase agrees with 
expectations of deconfinement, and the ``deconfinement'' transition takes 
place at the Hagedorn temperature, but eqs. 
(\ref{FreeE},\ref{FreeE<}) still do not represent a 
satisfactory state of affairs. The large \( N \) approximation of 
eq. (\ref{NSingletPartition}) does not give Hagedorn behaviour. 

The resolution to this puzzle is our main message. The Hagedorn and the deconfinement 
transitions can be identified, but one has to go to next order in 
the large \( N \) expansion to see the Hagedorn behaviour. This 
scenario was sketched for string theory in \cite{Atick:1988si}, but 
here we can see concretely the natural appearance of Hagedorn behaviour in large 
\( N \) field theory. (The possibility of such behaviour was discussed 
already by Thorn in \cite{Thorn:1981iv}.) Actually, 
although eq. (\ref{FreeE<}) results from the leading estimate of eq. 
(\ref{NSingletPartition}) they are of subleading order in \( 1/N \) and do 
not give the full \( N^0 \) contribution to thermodynamics. Fluctuations 
in the eigenvalue density give contributions of the same order and should 
also be taken into account. Below \( T_H \) this is remarkably easy (and 
above we do not have to do the calculation, if the goal is just to find 
Hagedorn thermodynamics).

The integral over eigenvalue-density fluctuations can only diverge at \( 
T_H \) because of the contribution from the integral over fluctuations in 
the eigen-directions corresponding to the eigenvalues that become 
degenerate with the low temperature solution at \( T_H \). The integral 
over these fluctuations given in eq. (\ref{DegenerateSolutions}) yields 
the correction factor
\begin{equation}
\sim{1 \over {\left( {1-\zeta(x)} \right)}}
\left( {1-e^{-N^2\left( {1-\zeta(x)} \right)}} \right),\quad \zeta(x)<1
	\label{NSingletPartition1}
\end{equation}
to the partion function, which diverges at the Hagedorn transition if the 
large \( N \) limit is 
taken before the approach to \( T_H \). Otherwise, the approach to the 
``deconfined'' phase is fast but smooth. This is just how the integrals 
tell us that an infinite number of degrees of freedom are needed for a 
true phase transition, at finite volume an infinite \( N \) is needed. 
With this reservation complete agreement is found with eq. 
(\ref{CanonicalHagedorn}), which of course was derived assuming an 
infinite \( N \).  

\section{\( Z_N \) symmetry}
\label{Z}

We have only associated the two phases of our model with confined and 
deconfined phases on the basis of the \( N \) dependence of the free 
energy. This is reasonable, but it should be compared to other criteria. 
A Polyakov loop, i.e. a Wilson loop around the compact imaginary time 
direction, is an order parameter for deconfinement. \( Z_N 
\) symmetry (the centre of the \( SU(N) \) symmetry) acts by 
multiplication by a root of unity on the Polyakov loop. Thus the loop has 
a vanishing expectation value in the confined phase, which enjoys unbroken \( Z_N 
\) symmetry. In the 
deconfined phase it gets a non-zero vacuum expectation value, due to 
spontaneous breaking of \( Z_N \) symmetry. Using the AdS/CFT 
correspondence the Polyakov loop has been studied at strong coupling 
in \cite{Witten:1998zw,Aharony:1998qu}. Our simple formalism is not 
well suited to a calculation of the loop, 
but there are alternatives. 

In our fluctuation calculation unbroken \( 
Z_N \) symmetry is 
essential to get agreement with the Hagedorn calculation in the confined 
phase. The symmetry appears in the integration over \( \alpha 
\)-translations, which results in a two-dimensional integral over 
fluctuations, giving the correct degree of divergence. On the high 
temperature side we have no similar check on the calculation. The best we 
can do is to judge from the partition sum itself how \( Z_N \) symmetry 
is represented. The eigenvalue densities (\ref{EigenvalueD}) clearly 
break \( \alpha \)-translation symmetry, so the issue is if the partition 
sum should be thought of as a sum over all possible translated minima of 
the effective action and fluctuations around them, or if a single minimum 
is selected. The simple answer is that there is no spontaneous breaking of 
symmetry in finite volume, even if \( N \) is large. A more satisfactory 
answer is that the high temperature limit, which in the conformal \( 
{\cal N}=4 \) SYM theory is the same as the infinite volume limit, 
eventually leads to the 
support \( [2\pi n/N +\alpha_{c}, 2\pi n/N -\alpha_{c}] \) of the 
equilibrium solutions for the eigenvalue densities (\ref{EigenvalueD}) 
narrowing so much that there is no overlap between densities 
related by \( Z_N \) transformations (different \( n\) from the \(\delta 
\) function in (\ref{NSingletPartition})). Even if the finite volume sum 
always is \( Z_N \) symmetric, this behaviour allows for a decoupling of 
fluctuations around different \( Z_N \)-related minima, which has to 
happen when \( Z_N \) is broken spontaneously. Since the high temperature 
phase is thus smoothly connected to a phase with spontaneously broken \( 
Z_N \), it makes sense to describe it as deconfined.

\section{AdS/CFT motivated speculations}
\label{AdS/CFT}

There is as yet no good test of the AdS/CFT correspondence 
\cite{Maldacena:1997re,Gubser:1998bc,Witten:1998qj,Aharony:1999ti} at weak 
coupling, so we cannot make strong predictions about gravitational
effects. We can however try to use the AdS/CFT dictionary, first to
compare with strong coupling results obtained via the correspondence, and 
second to find out what it tells us about a dual gravitational theory, 
assuming that the correspondence holds also at small 't Hooft coupling \( g^2 
N \), which means small string tension relative to the AdS curvature. 

At strong coupling Witten \cite{Witten:1998qj,Witten:1998zw} has used 
the AdS/CFT 
correspondence to discuss 
a large \( N \) phase transition at finite temperature. At leading order 
in large \( N \) he finds a high temperature phase with free energy 
proportional to \( N^2 \) and a low temperature phase where the 
(temperature dependent part of) the \( N^2 \) term in the free energy vanishes.
This precisely the behaviour we observe at weak coupling. It seems that 
thermodynamics on \( S^3 \) can have a smooth dependence on the coupling. 
In the flat case there is a debate \cite{Li:1998kd,Gubser:1998nz,Fotopoulos:1998es,
Vazquez-Mozo:1999ic,Kim:1999sg,Nieto:1999kc}  about whether there is a phase 
transition in the coupling or not,
and although 
most explicit calculations at extreme temperatures indicate a smooth 
behavior, the issue does not 
seem to be settled. 

Witten's result 
is obtained by associating the gauge theory behaviour with a 
gravitational phase transition between an AdS heat bath and an AdS black 
hole, found by Hawking and Page \cite{Hawking:1983dh}. It is of course tempting to 
identify the Yang-Mills thermodynamics we have found with a similar black 
hole transition on the gravitational side, but now in a theory of 
tensionless strings (zero 't Hooft coupling). 
Interpreted in gravitational language the \( N^2 \) 
dependence of the free energy translates to the ordinary \( 1/G_N \) 
dependence of a 
black hole free energy. The entropy would give an effective 
measure of the horizon area of this exotic black hole.

The \( N^0 \) corrections 
correspond to Hagedorn behaviour of tensionless strings\footnote{Their spectra 
are governed the curvature scale rather than the string scale.}, and beyond the 
Hagedorn transition we have a black hole equilibrium state. Such a phase 
diagram have been proposed for non-zero string tension \cite{Abel:1999rq}, by 
arguments based on a black hole/string correspondence principle 
\cite{Horowitz:1997nw}. Our explicit calculation is done at vanishing gauge 
coupling, but since \( N^0 \) corrections in the low temperature phase 
ought to be present independently of the coupling, this mechanism could 
give a general confirmation of the relation between Hagedorn spectra and 
black holes. It might seem strange that a string Hagedorn transition 
could occur at a scale governed by the AdS curvature rather than the string 
tension, but most of the important part of the Hagedorn spectrum consists 
of large strings, which should be more sensitive to curvature than to the 
tension. A complete picture might have to await the quantization of 
strings in \( \mathrm{AdS_5 \times S^5} \) with background Ramond-Ramond fields.

\section{Conclusions}
\label{Conclusions}

We have noted a general mechanism in large \( N \) theories, which can 
be important whenever the \( N^2 \) contribution to the free energy 
vanishes on one side of a phase transition. On this side, fluctuation can 
give an \( N^0 \) contribution which diverges at the transition. We have 
done this calculation explicitly for free \( {\cal N}=4 \) Super 
Yang-Mills theory on a three-sphere, and for quite a general class of toy 
models. The large \( N \) phase transition we find is a kind of 
deconfinement transition, in which Hagedorn behaviour in the confined 
phase is identified as a low-temperature precursor of deconfinement. 
This new mechanism appears to be sufficiently general to potentially 
apply to more physical theories. 

For asymptotically free gauge theories on \( S^3 \) we expect a 
similar behaviour in the limit of small radius since the coupling 
should vanish in this limit. Witten's results on the strong 't Hooft 
coupling limit of the \( {\cal N}=4 \) theory suggests that vanishing 
coupling may not be necessary for this kind of phase transition. Still, 
the large volume limit of the thermodynamics of asymptotically 
free gauge theories should be qualitatively different since it will 
depend on a scale \( \Lambda \) which is not present for \( {\cal N}=4 \).
It is obviously of great interest investigate this 
case\footnote{Supergravity solutions that could correspond to finite 
temperature non-conformally invariant field theories have recently 
been studied in \cite{Nojiri:1999uh}.}.

Assuming the AdS/CFT correspondence is valid also for small or vanishing 
't Hooft coupling, the transition has the likes of a Hawking-Page 
transition to AdS black holes, although in a string theory with vanishing 
tension. The \( N^0 \) correction we have calculated should then 
correspond to a Hagedorn contribution from an ideal gas of these strings.

\bigskip

\begin{flushleft}
This work was stimulated by conversations with U.\ Danielsson and H.\ 
Hansson, who are gratefully acknowledged. I also wish to thank J.\ 
Grundberg, U.\ Lindstr\"om and B.\ Nilsson for useful remarks. This work 
was financed by 
the Swedish Science Research Council.
\end{flushleft}

\end{document}